\documentclass[12pt]{article}
\usepackage[table]{xcolor}
\usepackage{latexsym,amssymb,amsmath,amsthm,amsfonts}
\usepackage[cmtip,arrow]{xy}
\usepackage{pb-diagram,pb-xy}
\usepackage{graphicx}
\usepackage{tikz}
\usepackage{multirow}
\usepackage{enumerate}
\usepackage{framed}

\begin{document}
\title{Analyzing Collaborative Forecast and Response Networks}
\author{Burcu Ayd{\i}n \and J.S. Marron}
\date{}
\maketitle
\begin{abstract}
Collaborative forecasting involves exchanging information on how much of an item will be needed by a buyer and how much can be supplied by a seller or manufacturer in a supply chain. This exchange allows  parties to plan their operations based on the needs and limitations of their supply chain partner. The success of this system critically depends on the healthy flow of information. This paper focuses on methods to easily analyze and visualize this process. To understand how the information travels on this network and how parties react to new information from their partners, this paper proposes a Gaussian Graphical Model based method, and finds certain inefficiencies in the system. To simplify and better understand the update structure, a Continuum Canonical Correlation based method is proposed. The analytical tools introduced in this article are implemented as a part of a forecasting solution software developed to aid the forecasting practice of a large company.
\end{abstract}
\section{Introduction} \label{Introduction}
Despite many efficiency gains achieved through investments in hub networks, information systems and data infrastructures by supply chain partners, forecast collaboration remains at the top of the list of obstacles to achieving supply chain goals. (See Gartner (2010)).

A recent study finds that reductions in forecast errors of the order of $1\%$ may be translated into inventory reductions in the order of $15$ to $20$ percent, and cycle service level and fill-rate improvements by approximately $1\%$. (See Syntetos et al (2010)).

Supply chain partners engage in information sharing to achieve better production planning and hence lower production costs for the supplier, and reduced stock-out costs and other risks for the buyer. The exchange allows the suppliers to form more clear expectations of the upcoming demand and plan the production schedules accordingly, resulting in a more precise supply flow. For the buyers, sharing of purchase plans in advance and providing regular updates results in reduced stock-out probability and inventory costs. For suppliers, signaling the upcoming production capacity limitations allow a smoother and more productive manufacturing process. The information exchange help the partners converge to a purchasing scheme through revisions. The forecast updates should take into account the information obtained from their partner’s forecasts issued in previous periods, as well as containing any new information that became available to the forecaster in that period. The updates of the forecasts from each party result in a structure called \emph{rolling horizon}.

Due to its nature of constant updates, rolling horizon forecasting is suitable for procurement engagements where collaboration between buyers and their suppliers is of importance. For example, for some types of materials, inventory for these parts are held by suppliers rather than the buyer, and the buyer issues detailed forecasts regarding future purchases, updated every period. (This is called \emph{Collaborative Inventory Management}, or CIM). This system allows the suppliers to form more clear expectations of the upcoming demand and plan the production schedules accordingly, resulting in a more precise supply flow. For the buyers, sharing of purchase plans in advance and providing regular updates results in reduced stock-out probability and inventory costs are eliminated.

In many settings, suppliers issue rolling horizon forecasts as well as a response to buyer’s purchase forecasts. These responses are communicated to buyers to give a signal on the production capacity of suppliers for the upcoming periods. They allow a better planning scheme for both the buyer and supplier, as the two are able to communicate about each other’s plans and capacities and converge to a better purchasing scheme through revisions made in each period. In fact, the forecast updates made by the buyer usually take into account the information obtained from responses to forecasts issued in previous periods.

Although the use of a rolling horizon forecasting scheme is very common due to its stated advantages, the rich and dynamic structure of the process brings difficulties in viewing and analyzing the vast data sets generated by it.


The aim of this paper is to introduce statistical methods specifically designed to simplify and understand rolling horizon forecasts. There are two main axes of variation in this data.


The first axis is variation of the information available for each time point. 
For any time period, there are many predictions issued by either side, all updates of each other. Understanding how these vary from period to period will provide insights into how the information shared evolves as the realization nears, as well as how parties react to the signals received from each other. These will be called the \emph{temporal trend} in the data.
To tease out an accurate picture of how information travels across the resulting information network, this paper proposes the use of a novel version of Gaussian Graphical Models (GGM).
GGM is a popular approach to analyzing gene networks in the literature.
The details are given in Section \ref{Temporal}.

The second axis is variation of the updates. 
Understanding the discrepancies between the actual realizations of each period and the predictions issued for it by each party is very important. Most supply chain collaborations come with contractual obligations that require a minimum amount of accuracy in the information shared by the partners. However, in the rolling horizon system, the information gets many updates as the realization period approaches. How to handle the many versions of this information in both tracking and contracting is not well understood. This challenge requires analyzing collective forecast information along the time horizon of interest. 

Section \ref{Progressive} of this paper proposes the use of Canonical Continuum Correlation Analysis (CCC) to analyze this variation while retaining a balance of within-forecast (within-response) variation and forecast-response correlation. This summary can be used for visual tracking of forecasts across periods, and can be an input for defining accuracy targets.

The two analytical tools described in this paper are developed as a part of a large forecast accuracy improvement initiative in Hewlett Packard. This initiative includes the development of a forecasting aid software, of which prototype codename is ANANSI. This tool is a web-based software that can be accessed by forecasters within the company. ANANSI includes dashboarding and real-time visualization capabilities, as well as a do-it-yourself analytics tool set. At the time this paper is written, the methods described in this paper are implemented, and scheduled to be rolled out to general use by the end of 2013.

The forecast accuracy improvement project, which this work is a part of, won the \textbf{APICS Corporate Excellence Award on Innovation} in 2012.

\section{Data Structure} \label{DataStructure}
The structure of the data generated by this process can be described as follows. At each period, forecasts regarding the upcoming $N$ periods are issued instead of only the immediate upcoming period. Therefore, at each time point $t$, the forecaster produces $N$ forecast numbers: $\{F_{t,t+1},F_{t,t+2},…,F_{t,t+N}\}$, where $F_{t,t+k}$ is a forecast issued at period $t$ predicting what will happen in period $t+k$. The difference between the time in which a forecast is made and the time for which the forecast is made, $k$, is called the \emph{lag} of that forecast number. In this process, the first $(N-1)$ numbers in the forecast series of each period can be considered as updates on the existing predictions made in previous periods, while the last forecast, $F_{t,t+N}$, is the first forecast being issued regarding the period $t+N$. At each period, the created forecasts are shared with the collaborating party, who in turn issues their own forecasts in the same structure as a response to the original forecasts. Note that the horizon length of responses may be different than those of the forecasts. When this is the case, the response horizon is denoted as $M$. These responses issued at period $t$ can be represented as $\{R_{t,t+1},R_{t,t+2},…,R_{t,t+M}\}$.

The analysis spans predictions for $T$ time periods. The forecast data generated in this process for purchases in the time interval $[t,T+t]$ can be collected into a matrix:
\[\textbf{F}=\left[F_1  \cdots  F_N \right]= \left[
                                   \begin{array}{c}
                                     F^t \\
                                     \vdots \\
                                     F^{t+T} \\
                                   \end{array}
                                 \right]=\left[
                                           \begin{array}{ccc}
                                             F_{t-1,t} & \cdots & F_{t-N,t} \\
                                             \vdots & \ddots & \vdots \\
                                             F_{t+T-1,t+T} & \cdots & F_{t+T-N,t+T} \\
                                           \end{array}
                                         \right]
                                 \]
In this matrix, each row $i$ represents  all forecasts made for the time period $i$. It is called the forecast \emph{dialogue vector} for $i$. Each forecast in the dialogue vector $i$ is an update on the next forecast on it. The dialogue vector of all the forecasts issued regarding period $i$ is denoted as  $F_i$.

Each column $j$ of the matrix represents all the forecasts that are issued $j$ periods before the period they are predicting. The column $F_j$ is the set of all forecasts with lag $j$, and is called an \emph{individual lag trend} with lag $j$.

The responses can be expressed in a similarly constructed $\textbf{R}$ matrix. Another possible input for analysis is the actual shipment vector $\textbf{S}$, containing the realized purchases. $S_i$ is the amount purchased at period $i$.

The nature of the process requires that $N\geq M$. For cases where $N>M$, we will take $R_N=\ldots=R_{M+1}=0$ for simplicity of presentation.


A natural way to directly understand several insightful analyses of this data structure is the concept of \emph{Object Oriented Data Analysis} (OODA). The terminology OODA was coined by Wang and Marron (2007). In its broadest sense, it refers to the thought process where data consists of objects of which population structure is statistically analyzed. Appropriate definition of atoms of analysis (objects) depends on the nature of the data as well as the aims of the analyst. Unlike the classical approach, the data are not necessarily seen as sets of numbers. Any input about a population of interest may be statistically analyzed in the OODA framework. Some examples of object definitions are vectors, shapes, images, or graphs. This flexibility allows statistical research on the increasingly rich and structured data that have become available through modern science and technology.

This paper focuses on two object definitions in the analysis of collaborative forecast data. Defining the dialogue vectors (rows) of forecast and response matrices to be the objects allows the focus to be on the variation among these rows. This object definition lends itself to clarifying the flow of information buried in the sequential updates of forecasts and responses, and thus allows direct analysis of the temporal trends in the data. The details of this approach are in Section \ref{Temporal}.

In Section \ref{Progressive}, the columns of forecast and response matrices, or the individual lag trends, are considered as the objects of the analysis. Modeling the interaction of these objects lends itself to an intuitive way of handling the updates and analyzing how the individual lag trends vary.
Their interaction can be considered in terms of their within-population variation, correlation, or a mixture of both.


\begin{figure}
\begin{center}
\includegraphics[natheight=4in,natwidth=6in,height=4in,width=6in]{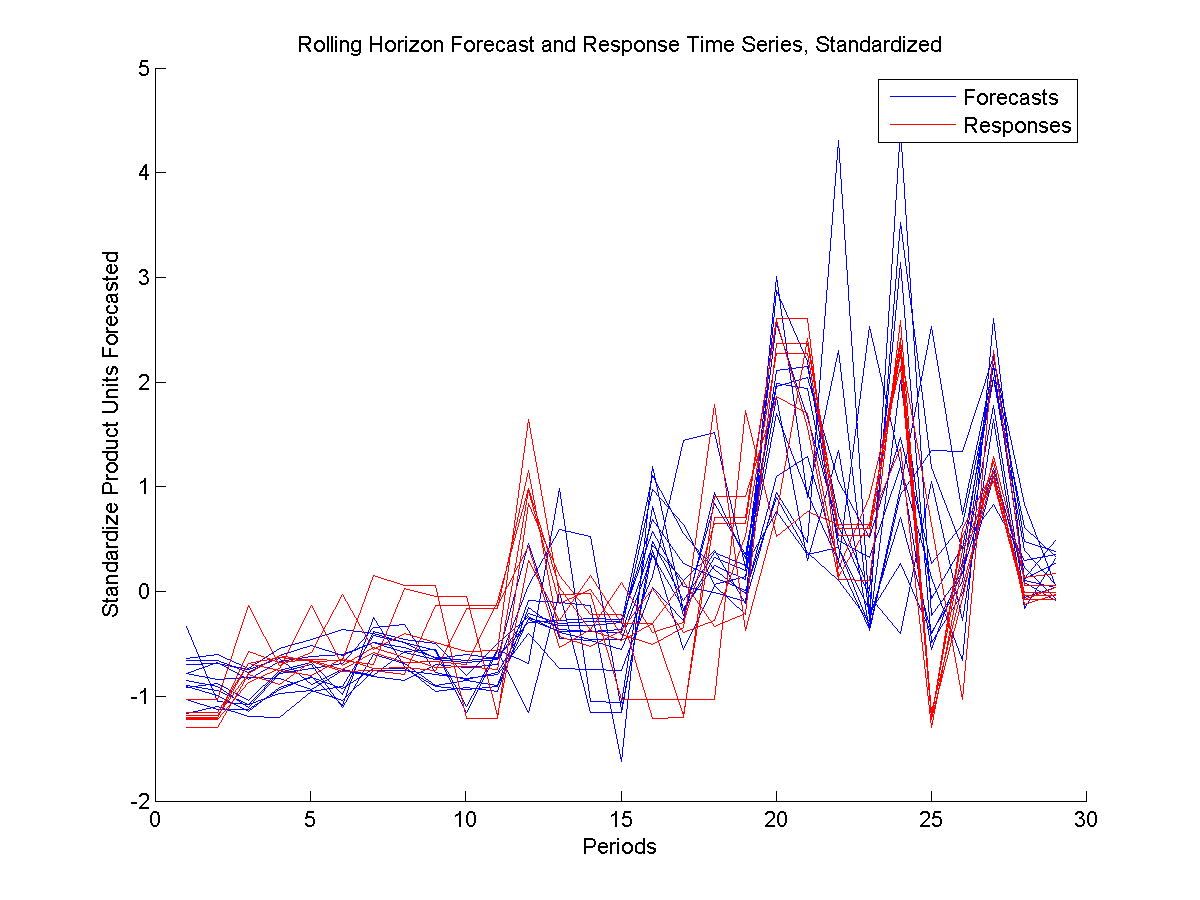}
\caption{An example set of forecast and response series groups. The X axis denotes the periods that each forecast is issued for. The Y axis is the standardized forecast amount. The blue series are forecasts, each series representing a different lag. Red series are responses.}\label{Fig1}
\end{center}
\end{figure}


\section{Temporal Trend: Rows as Data Objects}\label{Temporal}

The goal of this section is to build a model to understand the trend across the lags, and how the dialogue vectors vary. This goal requires focusing on the time structure of information issued by the partners for each shipment, and how they interact. This interaction can also be considered as flow of information across the lags and between the partners. A good way to model this interaction is to consider the structure of information exchange of the partners leading to each shipment as a network. In this network, each node will represent a forecast or response issued with  particular lag. These nodes will also be called \emph{events}.

In terms of the information contained, each event is the combination of the information propagated from the past events, and new information obtained in that period by the issuer. Understanding the interaction of the events is equivalent to correctly identifying which past events exerted influence on any given event. This will also enable the decomposition of the information contained in each event into what was propagated from the past, and the new information that entered this system with that event.
However, since past events  influence each other as well,  finding the correct source of each information component presents special challenges. Each piece of information that enters this system will most likely be incorporated into many events that come afterwards, echoing through the system. Therefore utilizing a local or pairwise approach is not appropriate.
A much better decomposition of this complex information flow through the network comes from the ideas of Partial Correlation and Gaussian Graphical Models.


\subsection{Background: Partial Correlation and GGM's}

Partial Correlation refers to the correlation between two random variables ($X$ and $Y$) when the effect of a set of external variables ($Z$) are removed from both of them. We will denote it as $C(X,Y|Z)$, where $C$ refers to correlation. Pearson's product-moment correlation coefficient is used as a measure of correlation. See Baba et al (2004) and Fisher (1924) for more information.

A good insight about partial correlation comes from a very simple example. Let the random variable $X$ represent ice cream sales over time and $Y$ sunscreen sales over time. It is reasonable to expect a high correlation between these two variables. However, this high correlation does not necessarily mean that one of these events is causing the other. In fact, if a variable expressing the season of the year ($Z$) is introduced, a most likely result will be that $C(X,Y|Z)$ is very small. This means that both $X$ and $Y$ (ice cream and sunscreen sales) are simply reacting to a third variable (summer). When the effect of summer is accounted for, $X$ and $Y$ are not related any more.

While the \emph{theoretical} partial correlation refers to the theoretical relationship between random variables, the \emph{empirical} partial correlation is a number estimated using observed samples from these random variables. Empirical partial correlation is an estimator for the theoretical partial correlation.

One way to find the  partial correlation between $X$ and $Y$, given $Z$, is to use linear regression:

\[X=Z\alpha+\varepsilon_{X|Z} , Y=Z\beta+\varepsilon_{Y|Z}\]

These regressions allow factoring  $Z$ out of both $X$ and $Y$. The correlation between the residuals $\varepsilon_{X|Z}$ and $\varepsilon_{Y|Z}$ can then be used to find the empirical  partial correlation: $C(X,Y|Z)=C(\varepsilon_{X|Z},\varepsilon_{Y|Z})$.

An equivalent approach to the partial correlation is to compute the empirical covariance matrix $S$ of the random variable set $[X,Y,Z]$. If $S$ is invertible, then the entry $S^{-1}_{1,1}$ (The 1,2 entry of the inverse covariance matrix) is proportional to the partial correlation between $X$ and $Y$, given $Z$. In the literature, the inverse of the covariance matrix is called the \emph{precision matrix}, or \emph{concentration matrix}, and is often denoted with $\theta$.

In a random variable set of arbitrary length, the entry $\theta_{ij}$ of the precision matrix is proportional to the partial correlation between variables $i$ and $j$ when the effect of all other variables in this system are removed from both them:

\[C_{ij}=-\frac{\theta_{ij}}{\theta_{ii}\theta_{jj}}\]

Partial correlation is a very powerful tool in isolating the interaction of two variables from the effects of the larger system they are in. Given a data set of many variables, the precision matrix of this system provides insights into the isolated interactions of any two variables. The larger entries in this matrix point to strong partial correlation, while smaller entries mean little or no correlation. A very important question at this point is to decide whether any small entry refers to weak but existing partial correlation, or if it is just a noise artifact. We approach this using \emph{sparse precision matrix } methods. These assume there are relatively few important underlying partial correlations, with the rest being practically zero. The goal is to identify and estimate the non-zero partial correlations. See Dempster (1972) and Cox and Wermuth (1996) for more information.

When the variables in such a system come from a multi-variate Gaussian distribution, the problem of finding a sparse precision matrix in such a system is solved through a \emph{Gaussian Graphical Model} (GGM). This is a well-researched area in the literature. See Yuan and Lin (2007) and the references therein for more information. Among other uses, they are extensively used to explore the interactions between genes in genetic studies. (Dobra et al (2004) ). These models are built to understand the interactions in a network of variables, where these variables are assumed to be following a multi-variate Gaussian distribution $N(\mu,\Sigma)$. A GGM network is represented by an undirected graph $G=(V,E)$, where $V$ is the set of vertices, and $E$ is the set of edges between the vertices. Each vertex corresponds to a random variable in this network. The edges describe the conditional independence between the random variables. An edge between two variables $X$ and $Y$ is absent whenever $X$ and $Y$ are essentially independent when conditioned on the other variables in the network. 

For variables $X$,$Y$,$Z$ following a multi-variate Gaussian distribution, the conditional independence of $X$ and $Y$ given $Z$ is equivalent to $C(X,Y|Z)$ being zero. Therefore finding a sparse precision matrix on this network leads to understanding the dependencies between the nodes. 

The GGM problem refers to estimating the existence or non-existence of the edges of the graph $G$, given a random sample from this network. Directly calculating the partial correlations from the samples will not give the desired results: because the input is a random sample, zero partial correlation will almost never be observed. However, values close to zero will indicate that the corresponding vertices are most probably independent, and the small non-zero partial correlation is just noise. Therefore, the challenge of GGM problem is to identify which entries of the precision matrix are actually zero with a certain level of confidence. This problem has been named the covariance selection problem, or the model selection problem in the Gaussian Graphical Network. 

For any GGM method, a parameter $\lambda$ is needed to control the tightness of the sparse model selection. A higher $\lambda$ will result in a network where many edges are pushed to zero. As $\lambda$ is lowered, the model selection will be more relaxed, and links with less certain conditional dependencies will appear. An appropriate $\lambda$ should be determined based on the data and the needs of the analysis. 

The literature contains many methods proposed to estimate the precision matrix given a desired tightness for model selection. In this work, the graphical lasso method proposed by Friedman, Hastie and Tibshirani in 2008 is used. This method is shown to be robust to data deviations, and it is computationally very fast.

The graphical lasso method solves the following problem for any given Gaussian graph $G$ and tightness parameter $\lambda$:

\[Maximize_{\hat{\theta}} \log|\hat{\theta}| -tr(S\hat{\theta}) - \lambda\|\hat{\theta}\|_1\]

Where $S$ denotes the empirical covariance matrix. In this formulation, the first two components represent the log likelihood function for the precision matrix $\theta$, and the last component is the linear shrinkage penalty to obtain sparseness.

Other approaches to the sparse GGM problem exist in literature. Some recent examples are explained in Banerjee et al (2008), Dobra and West (2004) and Meinshausen and B\"{u}hlmann (2006).

Some other studies in literature seek to relax the normality assumption of GGM's and therefore a provide more general framework. See Xue and Zou (2012) for a nonparanormal model, and Dobra and Lenkoski (2004) for a copula representation.

\subsection{GGM in an Information Flow Network}


The general GGM model does not consider a time dimension. The vertices do not represent events that happen in a sequence. To find the conditional dependence/independence between two vertices $X$ and $Y$, GGM controls for all the other random variables in the network.
Another consequence of the lack of a time-line is that the conditional dependence discovered between $X$ and $Y$ does not immediately provide information on causality. We do not know if  one variable may have caused the other; we only know there is dependence.

An information flow network like the subject of this paper, however, has an inherent time dimension. All events in the collaborative forecasting network happen in a sequence. This fact has two consequences.

First, when searching for the partial correlation between two events $X$ and $Y$, it is not appropriate to condition them on all the other events.
The appropriate action is to condition them only on the events of the past. The events that happened after both $X$ and $Y$ should not influence their dependence.

For deeper insight, suppose that this principal is not followed in finding the dependencies of this network. as an illustrative example, also suppose that the information flow network contains three consecutive events $X$, $Y$ and $Z$, and there is an information signal that entered the system with X, and then flowed to $Y$ and then $Z$. When investigating the flow between $X$ and $Y$, this information should be represented as a link between these two nodes. To apply the classical GGM methods to this network, the partial correlation between $X$ and $Y$ conditioned on $Z$ is calculated. Since $Z$ contains future information, this would result in inappropriate conditioning, where $X$ and $Y$ may appear to be conditionally independent.

Secondly, the time dimension allows us to construct our graph so that when a conditional dependence is discovered between $X$ and $Y$, the information contained in the earlier event ($X$) is propagated to the latter ($Y$), and not the other way around. Therefore we use a directed graph rather than the classical undirected graph of a GGM.

To accommodate the time dimension of our forecasting network, we first develop some appropriate notation, and then propose a scheme called \emph{Expanding Window GGM}, or \emph{EW GGM}.

Recall that $V$ denotes the set of nodes in this network. Let:

\[V_n=\{v_1,...,v_n\},\]

where $v_n$ refers to the $n^{th}$ event in the network. Notice that the indices indicate the sequence of events. Therefore $V_n$ refers to the set of first $n$ events. Furthermore, let $C^n$ denote the partial correlation matrix of the event set $V_n$.

\begin{figure}
\begin{center}
\includegraphics[natheight=4in,natwidth=6in,height=4in,width=6in]{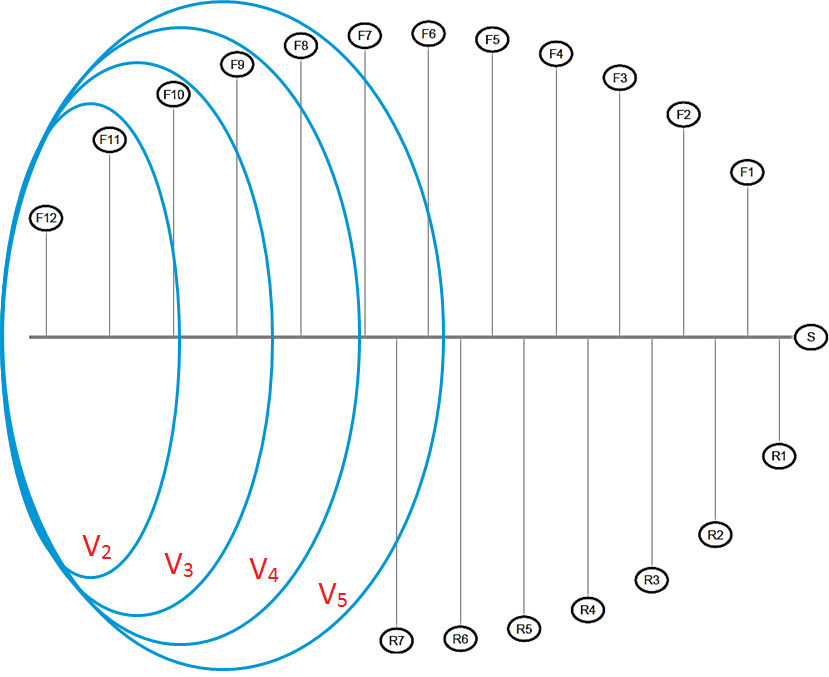}
\caption{The Expanding Windows idea illustrated. At each step, the algorithm focuses on the nodes that are in the current window, and solves the graphical lasso problem for this set. The resulting estimated precision matrix determines the incoming arrows to the rightmost node of the current window. For clarity of presentation, only the first $4$ windows are shown.}\label{ExpWin}
\end{center}
\end{figure}

In the information flow network, the partial correlation between any two nodes $i,j$ conditioned on all the events that happened before $i,j$ or between $i$ and $j$, is of interest. Let the matrix of all desired partial correlations be denoted by $C'$.
The $i,j^{th}$ entry of $C'$ is:

\[C'_{ij}=C(v_i,v_j|v_k \in V_m, m=max(i,j), k\neq i, k\neq j)\]

Then:

\[C'_{ij}=C^m_{ij}, m=max(i,j)\]

Therefore:

\[C'=\left[ C^1_{1}, \ldots, C^n_{n}\right], \]

where  $C^i_i$ is the column $i$ of the partial correlation matrix $C^i$, as defined above.

The problem of finding a sparse approximation of the $C'$ matrix is a difficult one. Note that, a simpler problem of finding a sparse precision matrix for the classical GGM problem was solved in a computationally efficient manner only recently. In this paper, instead of working towards an optimal solution, we will propose a heuristic approach that provides a robust approximation to the desired $C'$ matrix. We will also take advantage of the fast solution that is available for the classical GGM problem.

The Expanding Window GGM method relies upon providing sparse approximations for the columns $C^i_i$ through the sparse precision matrices $\theta^i$'s, calculated using the graphical lasso method for all $V_i$'s in the network. An illustration of the expanding windows of this idea is given in Figure \ref{ExpWin}.

Letting $S_k$ denote the empirical covariance matrix of the event set $V_k$, this heuristic method can be summarized as follows:

\begin{framed}
For a given ordered node set $V_n$ and a tightness parameter $\lambda$:\\

\ \ \ \ \ \ For $k$ from $2$ to $n$:
\[\hat{\theta}^k = \arg\max \log|\theta| -tr(S_k\theta) - \lambda\|\theta\|_1\]
\[C^k_{ij}=-\frac{\hat{\theta}^k_{ij}}{\hat{\theta}_{ii}^k\hat{\theta}_{jj}^k} \ \ \ \forall i,j \]

\ \ \ \ \ \ End

\ \ \ \ \ \ Set:
\[\hat{C'}=\left[ \hat{C}^1_{1}, \ldots, \hat{C}^n_{n}\right] \]
\end{framed}

%

\subsection{Information Decomposition}\label{decomposition}

In this system, each event is a combination of information that is available from the past periods, and the new information obtained in that period. The EW GGM method of the previous section provides a way to determine which past events contributed information to any given event. The next step is to determine the proportion of these contributions. Although the magnitudes of partial correlations expressed in $\hat{C}'$ give an indication of the size of the contributions, another step is needed for a precise decomposition.

Going back to the notation of Section \ref{Progressive}, the events consist of forecasts (F), responses (R) and a shipment event (S). We assume that the information in each event can be decomposed linearly.

\begin{figure}
\begin{center}
\includegraphics[natheight=3in,natwidth=6in,height=3in,width=6in]{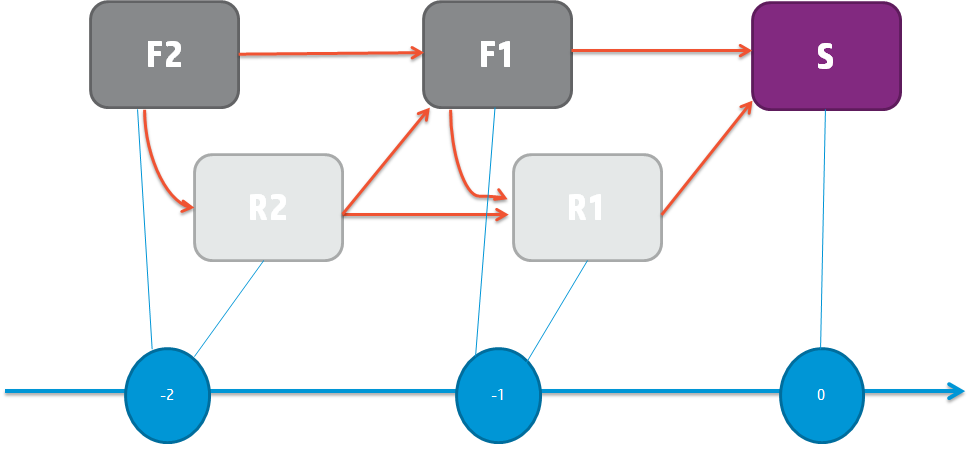}
\caption{The design of the collaborative forecasting system. Forecasts are dark grey, responses are light grey and shipment node is purple. The thick blue line is the time-line. The blue nodes represent the weeks before the actual shipment will occur. Red lines indicate influence, and thin blue lines connect each event to the time-line, indicating when the event occurs. Each event is supposed to incorporate  inputs from only the most recent forecast and response.}\label{ExpSys}
\end{center}
\end{figure}

Figure \ref{ExpSys} presents the intended design of the rolling horizon forecast collaboration system. Every forecast and response should include the signals that entered the system before it. Furthermore, we will also investigate the Markovian property.  This corresponds to the situation where the effect of the past on $F_i$ is fully captured by the events $F_{i+1}$ and $R_{i+1}$, and any previous event does not carry additional information. Therefore, the effect of, say, $F_{i+2}$ on $F_i$ should be close to zero when the effects of $R_{i+2}$, $F_{i+1}$ and $R_{i+1}$ are accounted for through conditioning.

Collaborative forecasting is intended to be Markovian, so it is interesting to investigate this property for the actual forecasting data. The EW GGM method of the previous section provides the actual links that are observed in this network. Based on the linear decomposition assumption, the following equations hold:

For lag $N$:
\[F_N=\varepsilon_N^F\]
\[R_N=\beta_N^RF_N+\varepsilon_N^R\]
For lag $i=1…(N-1)$:
\[F_i=\alpha_i^FR_{i+1:N}+\beta_i^FF_{i+1:N}+\varepsilon_i^F\]
\[R_i=\alpha_i^RR_{i+1:N}+\beta_i^RF_{i:N}+\varepsilon_i^R\]
Where $F_{i:N} =[F_i,\ldots,F_N]$.

For S:
\[S=\alpha^SR+\beta^SF+\varepsilon^S\]

The coefficients in this system are calculated using regression.

Recall that the empirical partial correlation matrix of the network has zero entries for events that are partially independent. In the decomposition step, the coefficients that correspond to a partially independent pair of events are set to zero. In other words, the information in each event is decomposed using only the past events from which an incoming arrow to this event exists. Therefore the EW GGM method serves as a model selection tool in decomposition. Note that the sparsity of the estimated partial correlation matrix is controlled by the parameter $\lambda$. Higher values of $\lambda$ will result in tighter model selection in this step.

The non-zero $\alpha$ and $\beta$'s obtained through regression correspond to the magnitude of the influence of past events, while the $\varepsilon$'s correspond to the new information and noise.

\subsection{Data Transformation}

Standardizing the input data is a common practice in statistics. It prevents scaling issues from dominating the results, and allows the correct detection of correlations. Another common practice is to test for the normality of input data, and apply a monotone transformation to the data if its distribution is noticeably different from normal. The transformation allows the analysts to take advantage of the attractive properties of the normal distribution. Moreover, hypothesis testing (checking whether an observed relationship is actually there or not) can be reliably done with appropriately transformed input.

Our method assumes Gaussian input data. Unfortunately, there tends to be strong skewness in collaborative data. Therefore we use recommend use of the Box-Cox transformation of Box and Cox (1964):
\[ y_i^{(\gamma)}=  \left\{
                        \begin{array}{ll}
                          \frac{y_i^\gamma-1}{\gamma}, & \text{if \ \ } \gamma \neq 0; \\
                          \log(y_i), & \text{if \ \ } \gamma = 0.
                        \end{array}
                      \right.
\]

\begin{figure}
\begin{center}
\includegraphics[natheight=5in,natwidth=6in,height=5in,width=6in]{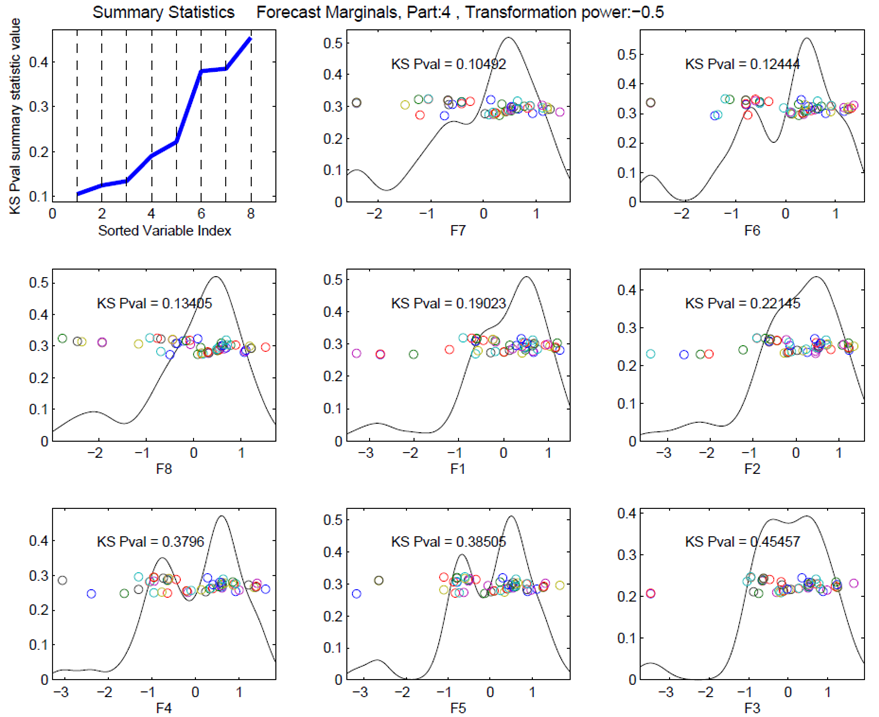}
\caption{The summary statistics and empirical distribution plots for the box-cox transformation of an example part with a $\gamma=-0.5$. The plot includes last $7$ lags of the forecasts, and a sorted variable index. In each plot, the p-value resulting from the Kolmogorov-Smirnov normality test is given. These show that normality cannot be rejected for any of the given forecast lags for this $\gamma$. }\label{SumStat}
\end{center}
\end{figure}

The proper $\gamma$ is found through exploration. In our data set, a value of $\gamma=-0.5$ seems to achieve reasonably Gaussian data for some representative parts. The plots of empirical distributions obtained using this value of $\gamma$ for one example is shown in Figure \ref{SumStat}.

After normalization, as the second step, we standardize the transformed input. Notice that the equations given in the previous subsection do not contain a constant term, because all columns are scaled to have zero mean.

\subsection{Visualization}

\begin{figure}
\begin{center}
\includegraphics[natheight=4in,natwidth=6in,height=4in,width=6in]{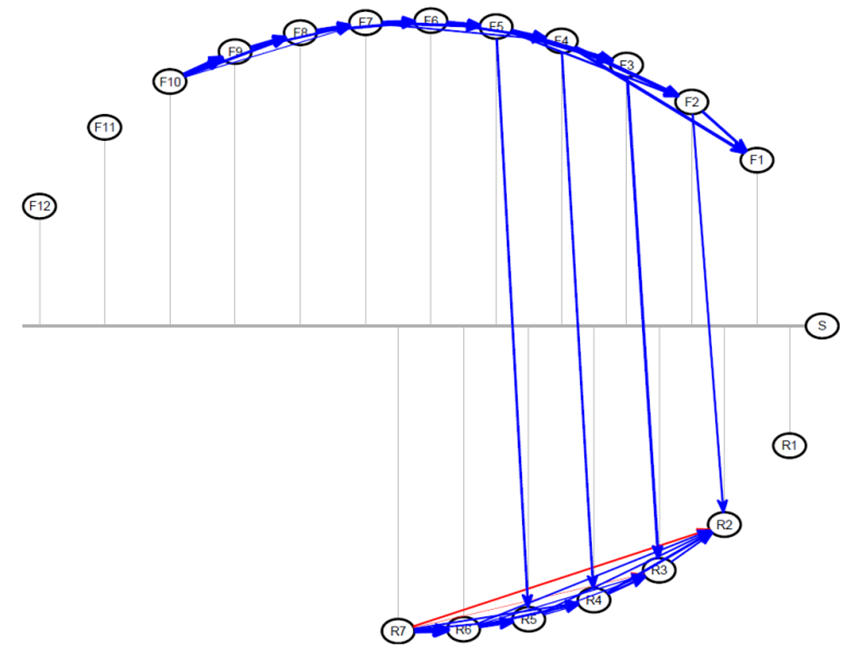}
\caption{An example forecast-response network. The arrows indicate which past events influence any given event, determined by the EW GGM algorithm. The thickness of the arrows are proportional to their regression parameters obtained through the decomposition step. Color of the arrows indicate the sign of the parameter: blue is positive and red in negative.}\label{network}
\end{center}
\end{figure}

We propose a new visualization technique to represent the decomposition of the information flow network. Figure \ref{network} shows an example forecasting network illustrated in this fashion.

In Figure \ref{network}, the grey horizontal line represents the time-line. Each of the forecast, response and shipment events is shown as a node. The $X$ coordinate of the position of each node indicates the time point that the event happened on the time-line. The vertical grey lines from the nodes can be used to identify when each event happened. The $Y$ coordinates of the nodes are determined such that the nodes are arranged on an ellipse. This allows us to present the edges with least amount of overlap. The forecasts are arranged on the top hemisphere of the ellipse and responses are on the lower hemisphere to enable quick visual inspection of forecast-response interaction. The shipment node, if it exists, sits on the rightmost end of the time-line.

An edge from node $i$ to node $j$ is drawn if the EW GGM method indicates that these events are conditionally dependent. The directions of these edges are always from left to right (from past to more recent). The color and thickness are determined by the decomposition regression. For each node, the color of the incoming edge is blue if the corresponding parameter in that node’s regression is positive, and it is red if the parameter is negative. The thickness of the arrow is proportional to the magnitude of the parameter, indicating which past events exerted large and small influences.

\subsection{Results}

The methods described are applied to the data set coming from the collaborative forecasting practice between HP and a microprocessor provider. The data contains forecasts with up to $12$ lags, and responses with up to $7$ lags.

In this section, we will illustrate the use of this tool on an example unit from this data set. The selected unit is a microprocessor used in North America region. The unit is selected due to its strategic importance and high volume. The observations obtained from this unit are typical among a larger collection that was considered. Thus the similar results for other representative units are presented in supplemental files and not in the main text.

The first decision the user will make is the choice of the model selection parameter. The forecast analytics software tool allows users to interactively change this parameter and review results in real time. For the purposes of this paper, the networks obtained by changing the model selection parameter in the range $[0.7,1]$ are provided in a supplementary document. A quick exploration of this document using the  up and down arrow buttons allows the user to observe the effect on the application. In our experience, the parameter range around $0.7-0.8$ gave too little model selection, i.e. way too many very small links were shown while  $0.9-1.0$ was too tight, erasing most of the links. Approximately $0.8-0.9$ gave the most reasonable range. In fact, $0.8$ seems to be a reasonable value for most of the units in the data set. The forecast-response network of our example part for this parameter value is in Figure \ref{network}.

Recall that we expect to see the Markovian property in this network. If it is there, past influence on each event should be fully captured by the most recent forecast and response. It would visually present itself by short, fat arrows coming in to each event only from the most recent events. In Figure \ref{network}, we see that this property is mostly there for within forecasts and within responses.

A striking observation in this graph is the lack of arrows going from response nodes to forecast nodes. This indicates that the buyer is not strongly using the information coming from the supplier when revising the forecasts. The adoption of the forecasts based on the supplier’s signals is an important goal of this setting, so our visualization provides an important diagnostic tool, which gives insight into how well this system is working in practice.

The next observation is, the forecasts are closely linked with each other, and the responses are closely linked with each other, as expected. On the other hand, the only influence between forecasts and responses appear from $F5$, $F4$, $F3$ and $F2$ to the responses $R5$, $R4$, $R3$ and $R2$, respectively. This makes sense, because for this particular microprocessor, the production delay is $4$ periods. Therefore, when deciding how much to produce $4$ periods ahead, the most recent demand forecast that the supplier has access to is $F5$. The supplier does not react to earlier forecasts because they are not recent at the time of production decision. The updates communicated in $F4$, $F3$ and $F2$ are also incorporated to supplier's response through supplier's manufacturing flexibility, buffer inventory and re-arrangement of product commitments. However, $F1$ carries a very late update, perhaps too late, so the supplier does not react to it.

We see that the events $F12$, $F11$, $R1$ and $S$ are not linked to any other event. $F12$ and $F11$ are the earliest events. It may be the case that it is yet too early to make meaningful projections for weekly purchase, and thus these nodes do not carry significantly meaningful information. Similarly $R1$ is the response right before the realization of the purchase. It is not meant to influence any forecasts. Knowing this, the responders do not seem to put meaningful information in it. Given that these nodes do not provide meaningful information, and they take effort to produce, it may be beneficial to remove them from the process to save forecasters' time. Thus our visualization tool provides a large amount of useful information for streamlining the forecast response system.

The fact that the event $S$, shipment, is not connected to any other event is an indication of forecast inaccuracy in the system. This may be the most important problem to fix in this system, since forecast inaccuracy against the actuals is the biggest driver of buffer inventory and stock out costs.

Finally, recall from Section \ref{decomposition} that the magnitude of any influence is communicated through the thickness and color of arrows. When it is more useful to know the exact values in the underlying decomposition, these will be given using a mouse-over action on the tool. As an example, the decomposition of the $F1$ event in this network is:

\[F_1=0.5F_2+0.4F_4+\varepsilon_1^F\]

This decomposition implies that $50\%$ of the information contained in $F1$ propagated from $F2$, $40\%$ came from $F4$, and $10\%$ is new information and noise.


\section{Individual Lag Trends: Columns as Data Objects} \label{Progressive}
The main goal of this approach is to understand trends across periods (along the X axis in Figure \ref{Fig1}), and how the individual lags (each curve) vary.

These trends have two main components. The first component is change in the forecast (response) information from period to period. The second component is about the relationship between the forecast and response groups. As a whole, responses are supposed to follow the forecasts as closely as supplier’s capabilities permit. Therefore, a high correlation is expected between the groups. On the other hand, the responses should also signal the availability situation of the suppliers, so responses fully mimicking forecasts are not realistic.

A robust and informative method to analyze the first component is \emph{Principal Component Analysis} (PCA). As a simple eigenvector-based multivariate analysis, PCA orthogonally transforms the variables of a mean-centered data set into linearly uncorrelated variables, called principal components. The first component carries the maximum amount of variance. The second component carries the maximum amount of remaining variance and it is orthogonal to the first component. See Jolliffe (2002) for more information on the method.
The first principal component for the forecasts (responses) can be viewed as the direction vector that goes through the cloud of individual lag trend objects while retaining the maximum amount of variation, or equivalently, minimizing the total sum of squared distances of these objects to this direction.

To examine the relationship between forecasts and responses, \emph{Canonical Correlation Analysis} (CCA) is a natural candidate. CCA finds two vectors with maximum correlation with each other, where one of these vectors is a linear combination of one group of series (forecasts) and the other is a linear combination of the other group (responses). These two vectors can be thought of as directions representative of their respective sources where the correlation information between them is as amplified as possible. These series show at which time periods the forecast and response groups are the most aligned and at which periods they diverge.

A very common problem plaguing CCA and many other similar methods is overfitting. A rule of thumb suggests having at least $10$ observations per variable to avoid overfitting. (See Hair et al (1998)). A typical data example for the forecasting problem this study focuses on, as seen  in Figure \ref{Fig1}, has $20$ data series (variables), and $30$ periods (observations). In this example, even if the two groups were completely uncorrelated, the CCA would show indications of correlation through over-fitting. 

Although the approach we propose in this paper is a remedy for overfitting, there are other solutions in literature. \emph{Regularized CCA} can be a way to tackle over-fitting. Regularization in the context of CCA was first proposed by Vinod (1976), then developed by Leurgans et al. (1993).

Another consideration is the need to balance and combine the two goals of considering variation and correlation. \emph{Partial Least Squares} (PLS) Analysis can be used to find summary direction vectors that maximize the covariance between forecast and response groups. This method is in some sense a midway between PCA and CCA.

Although PLS offers a convenient single method to account for both within group variation and inter-group correlation, it assigns weights which are in some sense equal on variation and correlation. This may not be suitable in every situation. In fact, in our current data set forecast variation is much larger than forecast-response correlation, therefore the PLS results are dominated by the variation. The PLS results are virtually the same as the PCA results.

To study the correlation more deeply, we adopt a relatively new method called \emph{Continuum Canonical Correlation} (CCC) Analysis.

Stone and Brooks (1990) proposed an umbrella formulation called \emph{Continuum Regression} which encompassed ordinary least squares, partial least squares and principal component regression methods. Through a single tuning parameter $\alpha \in [0,1]$, continuum regression includes these three methods as special cases where $\alpha$ is $0$, $0.5$ or $1$. The parameter $\alpha$ controls the trade off between the variance of the input data and covariance between the input and output data.

In 2007, Lee and Marron extended continuum regression ideas to analyze two sets of multi-variate data cases in the spirit of CCA. The new method, CCC, contains CCA, PLS and PCA as special cases, and seeks to find directional vectors in two sets of data while seeking to balance the inner-population variation of each data set with the inter-data set correlation. This balance is tuned using the parameter $\alpha$.

In the forecast-response setting of this paper, the two input data sets are $\textbf{F}$ and $\textbf{R}$. The core of CCC is based on finding vectors $F^*$ and $R^*$, where each of them is a linear combination of columns of $\textbf{F}$ and $\textbf{R}$, respectively:

\begin{equation} \label{ccc1}F^*=w_1*F_1+\cdots+w_N*F_N=\textbf{F}*\textbf{w}\end{equation}

\begin{equation}\label{ccc2}R^*=v_1*R_1+\cdots+v_N*R_N=\textbf{R}*\textbf{v}\end{equation}

\begin{equation}\label{ccc3}\|\textbf{w}\|=\|\textbf{v}\|=1\end{equation}

Where $\textbf{w}=[w_1 \cdots w_N ]^\prime$ and $\textbf{v}=[v_1 \cdots v_N]^\prime$. By requiring the weight vectors $\textbf{w}$ and $\textbf{v}$ to have norm one, $F^*$ and $R^*$ are ensured to be standardized linear combinations of $\textbf{F}$ and $\textbf{R}$.

The mathematical formulation to be solved to obtain the $\textbf{w}$ and $\textbf{v}$, depending on the user-defined $\alpha$, can be given as follows:

\begin{equation}
\begin{aligned}
& \underset{\textbf{w},\textbf{v},F^*,R^*}{\text{Maximize}} & & COV(F^*,R^*)^2 \left(VAR(F^* )VAR(R^* )\right)^{\frac{\alpha}{1-\alpha}-1} \\
& \text{Subject to:} & &  (\ref{ccc1}),(\ref{ccc2}) \text{ and } (\ref{ccc3}).
\end{aligned}
\end{equation}

Or equivalently:

\begin{equation}
\begin{aligned}
& \underset{\textbf{w},\textbf{v}}{\text{Maximize}} & & \left(\left(\textbf{Fw}\right)^\prime(\textbf{Rv})\right)^2 \left((\textbf{Fw})^\prime (\textbf{Fw})(\textbf{Rv})^\prime (\textbf{Rv})\right)^{\frac{\alpha}{1-\alpha}-1} \\
& \text{Subject to:} & &  (\ref{ccc3}).
\end{aligned}
\end{equation}
The vectors $F^*$ and $R^*$ contain summarized information on variance and correlation of the full $\textbf{F}$ and $\textbf{R}$ matrices. The choice of linear combination vectors $\textbf{w}$ and $\textbf{v}$ depends on the type of information the user is seeking, coded in the parameter $\alpha$.

$F^*$ and $R^*$ are time series of length $T$, just like the revision time series’ they are obtained from. They can be also considered as directions that summarize information contained in all the $N$ raw time series. The members of vectors $\textbf{w}$ and $\textbf{v}$ determine which raw time series will have greater (or less) weight in $F^*$ and $R^*$.

When $\alpha=0$, the objective function becomes $\left(\frac{COV(F^*,R^*)}{\sigma(F^*)\sigma(R^*)}\right)^2$, the square of Pearson's correlation coefficient, denoted as $\rho_{F^*,R^*}$. Maximizing $\rho_{F^*,R^*}^2$ under the given constraints is equivalent to maximizing $\rho_{F^*,R^*}$, and thus performing canonical correlation analysis. When $\alpha=0.5$, the objective function becomes $COV(F^*,R^*)^2$, of which maximization is equivalent to performing partial least squares analysis. At $\alpha=1$, the objective function is undefined, but as $\alpha\rightarrow 1$, the objective approaches the equivalent of maximizing $VAR(F^*)VAR(R^*)$. This problem can be decomposed to maximizing $VAR(F^*)$ and $VAR(R^*)$, or performing principal component analysis to $\textbf{F}$ and $\textbf{R}$.

Continuum Canonical Correlation provides an encompassing method of which all these three methods are special cases. The relative amounts of variance or correlation that will be emphasized in the summary vectors $F^*$ and $R^*$ is controlled by the tuning parameter $\alpha$. By continuously varying $\alpha$ between $0$ and $1$, a full spectrum of analyses between these methods can be performed. 

Lee and Marron (2007) propose an algorithm to find the optimum vectors $\textbf{w}$ and $\textbf{v}$ of this formulation. Alternatively, generic non-linear solvers may also be employed.

Different values of $\alpha$ between $0$ and $1$ enable the user to see how the summary changes depending on how much weight is given to either objective. Changing this value and interactively observing the change in the resulting outputs allows the user to understand the variation and correlation existing in the data set. Furthermore, the comparative relationship between these effects becomes apparent when objectives balancing these with different weights are examined.


\subsection{An Example}

In this section, we will illustrate the CCC method on a microprocessor part. In the ANANSI tool, the method is implemented such that the user can interactively change the $\alpha$ parameter and see how both the summary plot and the forecast-response scatter plot change. In the paper, we will provide these plots for the $\alpha$ values $0$, $0.5$ and $1$.

\begin{figure}
\begin{center}
\includegraphics[natheight=2in,natwidth=2.5in,height=2in,width=2.5in]{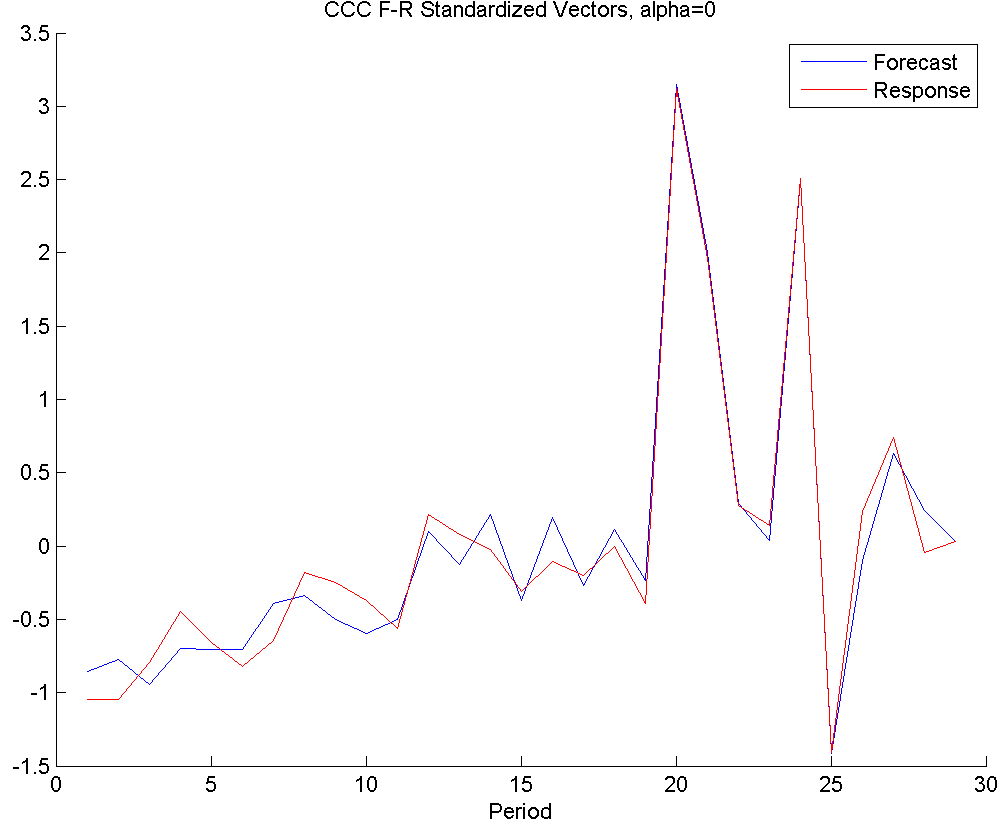} \includegraphics[natheight=2in,natwidth=2.5in,height=2in,width=2.5in]{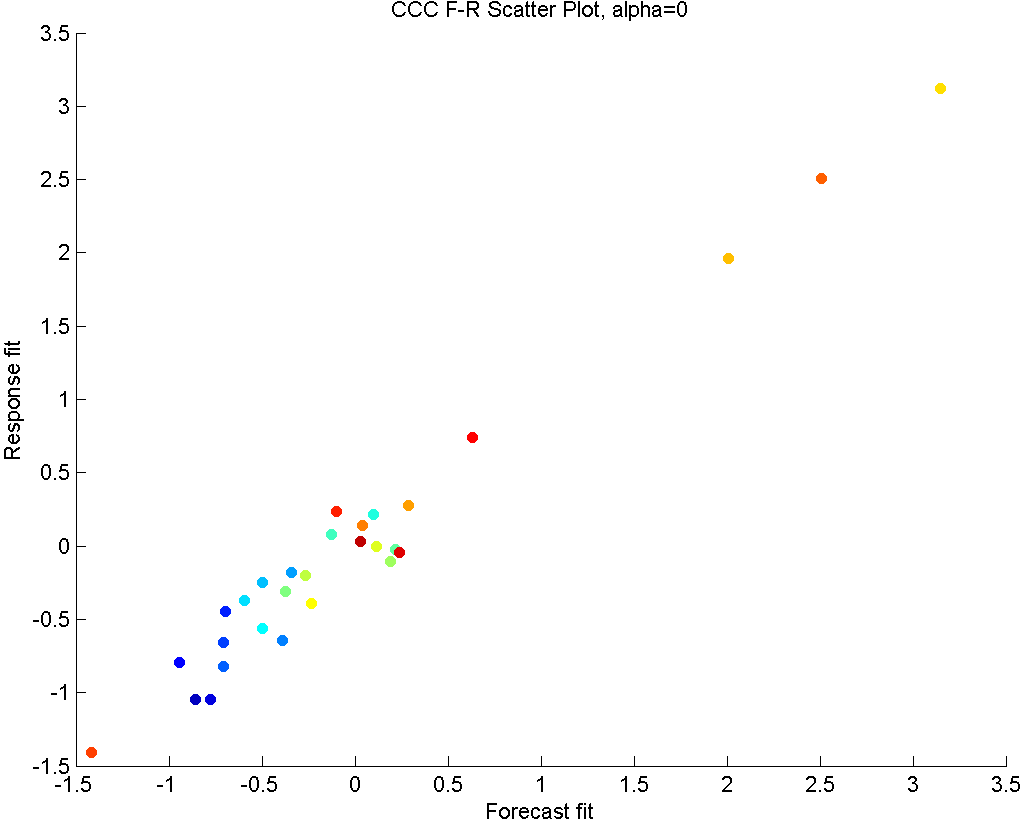}
\includegraphics[natheight=2in,natwidth=2.5in,height=2in,width=2.5in]{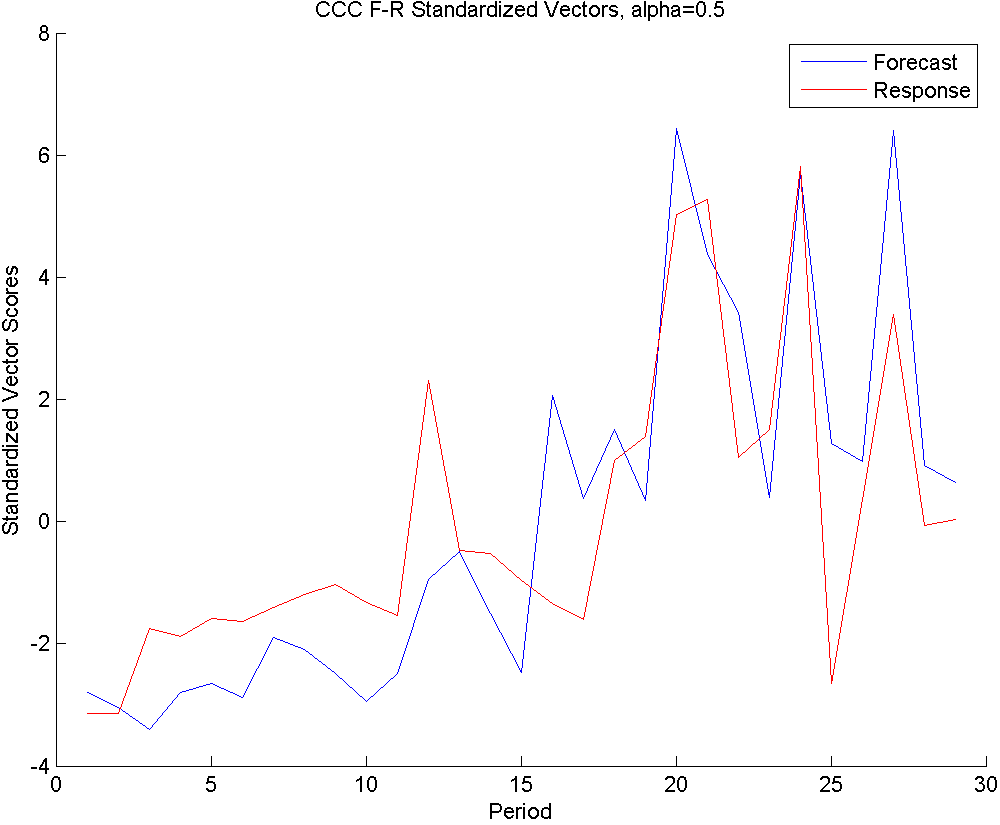} \includegraphics[natheight=2in,natwidth=2.5in,height=2in,width=2.5in]{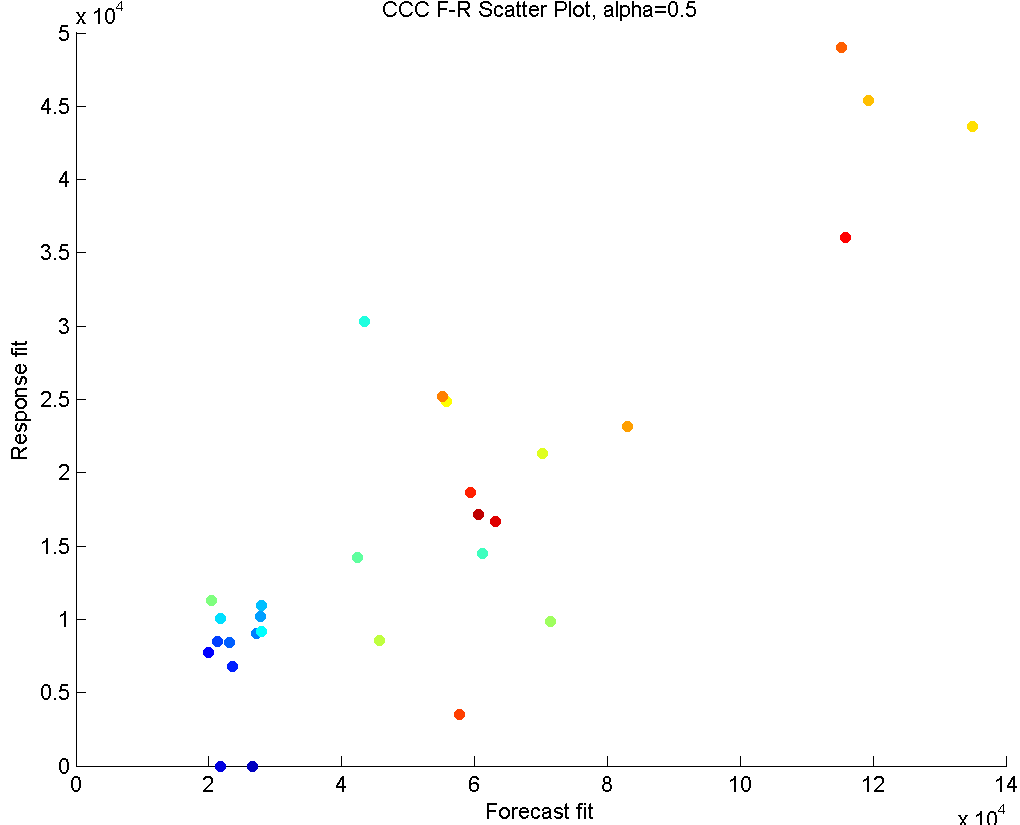}
\includegraphics[natheight=2in,natwidth=2.5in,height=2in,width=2.5in]{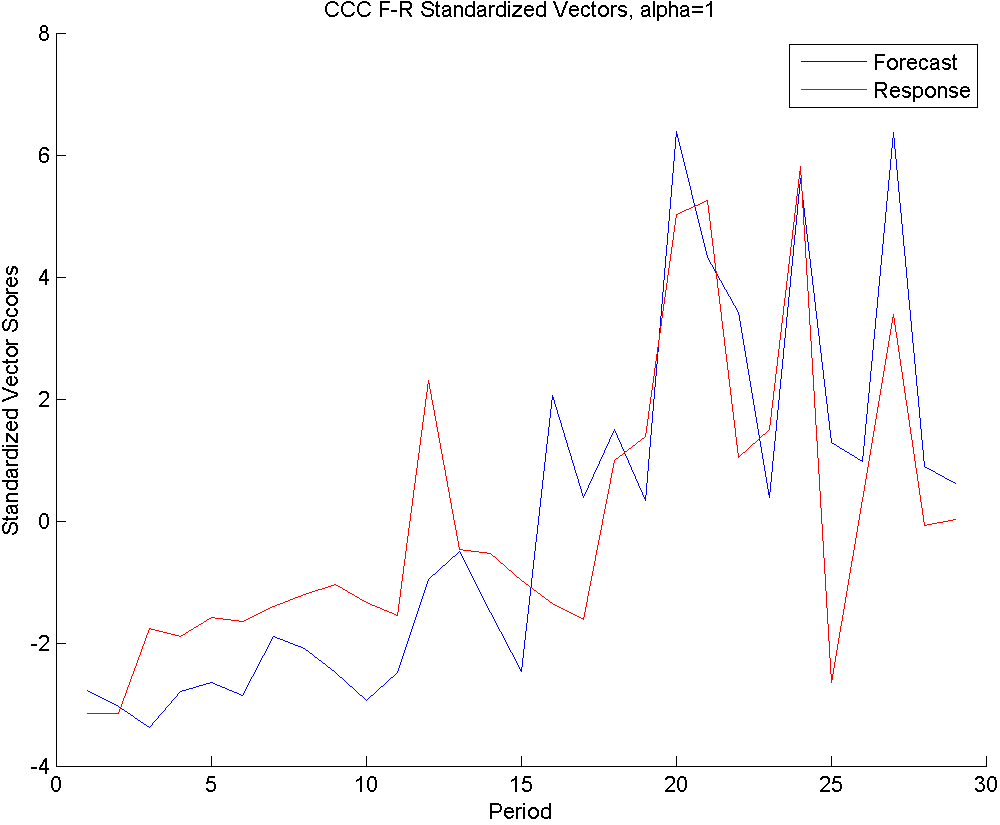} \includegraphics[natheight=2in,natwidth=2.5in,height=2in,width=2.5in]{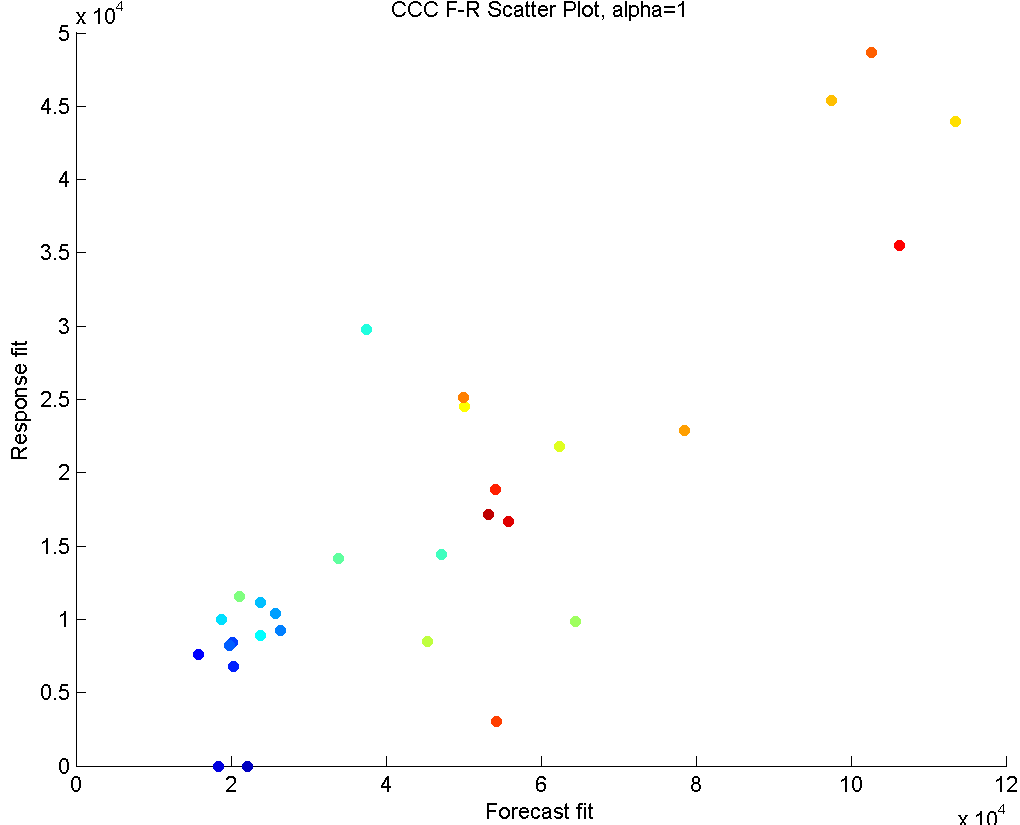}
\caption{On the left: The F* (blue) and R* (red) vectors found using CCC. On the right: Scatter plots where each dot is one period. The X coordinate of each dot is its score on the forecast vector and the Y coordinate is its score on the response vector. The colors range from blue to yellow to red where early periods are closer to the blue end and later periods are closer to the red end on the spectrum. The rows are graphs for $\alpha=0$, $0.5$ and $1$.}\label{alpha}
\end{center}
\end{figure}

In Figure \ref{alpha}, we see the results of applying CCC to this part.

The first row shows the plots with $\alpha=0$. This means the method only seeks correlation structure. On the left panel, we see that a very strong correlation exists in later periods, which indicates close collaboration and coordination between the parties. In earlier periods, although some correlation exists, it is not as strong. This is most probably due to the fact that in the initiation periods, supplier kept a certain level of inventory, but the buyer orders weren’t strong yet, so there wasn’t a strong need for close collaboration.

Similarly, the scatter plot on the right hand side indicates a very strong linear trend, with cooler colors (earlier periods) occupying the lower end of order sizes, and warmer colors (later periods) make up the higher order size periods. An outlier in this plot would indicate a period where the forecast and response values in all horizons were unrelated to each other, suggesting a possible communication problem or a maybe a disagreement. Such a period does not exist in the given example.

Even though some insights can be gained from these graphs, it is clear that both plots show an excessive amount of correlation, due to overfitting.

In the second row, we have $\alpha=0.5$. This option considers both the variance and correlation equally. We see that the summary vectors of forecasts and responses very closely follow their $\alpha=1$ vector summaries, which only considers the variance. This means that the within-sets variability for this data set is much stronger than the correlation effect, causing these vectors to feel the variance more strongly than the correlation. Note that this conclusion can only be made when $\alpha=0.5$ results are compared to the other two results.

In the third row, we have $\alpha=1$, which gives the summaries of forecasts of responses without taking into account the relationship between them. It only considers to retain the information existing within the forecasts (blue line), and separately, the information within the responses (red line).

The summary forecast and response series show a low level of volume in the first half of the periods, and a high and rather volatile forecasting pattern for the second half of the periods (Left panel). This suggests that this part was initiated around periods 1-10 with low volumes (and sufficient supplies), and the consumption picked up in the later periods (where supply predictions were slightly lower for certain periods). The trends of forecasts and responses are close to each other, indicating that a long term strong supply shortage is not in effect.
Although these summary vectors only consider the within-set variance, the second half of the periods show a large amount of correlation. This can also be seen in the scatter plot of the right panel. A linear trend in this plot would suggest that responses and forecasts are moving together. Like with the other $\alpha$ values, the blue-green hue dots crowd the lower left area of the scatter plot, while reds are closer to upper right. This tells us that high predictions exist in later periods, while the earlier periods have relatively less volume. Outliers from the linear trend would suggest a big shortage period (this would show up in lower right) or an excessive inventory period (this would show up in upper left). Our example does not possess strong outliers.

If a certain default value for $\alpha$ to be followed is desired, these plots make it clear that a good balance between the variance and correlation can be obtained somewhere between $\alpha=0$ and $\alpha=0.5$,  Interactively going through all the possible $\alpha$ values (not given here), we see that a value around $0.1$ provides an acceptable balance.

For future, it may also be beneficial to approach this method with a mechanism design perspective. A certain value of $\alpha$ acceptable to both buyer and supplier can be decided on to follow, and deviations and fluctuations defined over the summaries based on this $\alpha$ can be penalized. The effect of such an agreement on the forecast accuracy and collaboration, and the best values of $\alpha$ to achieve supply chain goals of interest are left to future research.

\bibliographystyle{abbrv}

\end{document}